\documentstyle[12pt]{article}
\if@twoside \oddsidemargin 21pt \evensidemargin 59pt \marginparwidth 85pt
\else \oddsidemargin 0pt \evensidemargin 0pt
\fi
\begin{document}

\title{A Mathematical Model for the Behavior of Pedestrians}

\author{Dirk Helbing \\II. Institut f\"{u}r Theoretische Physik\\
Universit\"{a}t Stuttgart}
\maketitle

\begin{abstract}
The movement of pedestrians is supposed to show certain regularities which can
be best described by an ``algorithm'' for the individual behavior and is easily
simulated on computers. This behavior is assumed to be determined by an
intended velocity, by several attractive and
repulsive effects and by fluctuations. 
The
movement of pedestrians is dependent on decisions, which have the purpose of
optimizing their behavior and can be explicitly modelled. Some interesting
applications of the model to real situations are given, especially to formation
of groups, behavior in queues, avoidance of collisions and selection
processes between behavioral alternatives.
\end{abstract}

{\small {\bf Key words:} pedestrians, movement, dynamics, motivation, conflicts,
decisions, field theory, groups, queues, avoidance, territory, 
selection, break of symmetry}

\section{Introduction}

Human behavior is based on
individual decisions. In building a mathematical model
for the movement of pedestrians, one has to assume that these decisions
are not completely random, but show
certain regularities instead. This assumption
may be justified, because decisions and therefore the behavior of pedestrians
will usually be determined by utility
maximization: A pedestrian wants to move in a most convenient
way, tries to minimize delays when having to avoid obstacles and other 
pedestrians, intends to take an optimal path and to walk with the minimal
velocity allowing to reach a destination at a certain time, etc. The
optimal behavior for a given situation can be derived by plausibility
considerations and will be used as a model for pedestrian movement.
Of course this optimal behavior is normally not thought about by an
individual, but by trial and error it has automatically learned to use
the most successful behavioral strategy, when being confronted with a 
standard situation 
(compare to sect. 3.2,(d)).
\par
Due to several reasons we cannot expect the model to be {\em exactly} valid.
Firstly an individual may find itself in a nonstandard 
situation. Secondly it probably
has not learned the optimal strategy yet. Thirdly sometimes emotional or other
reasons may lead to a suboptimal behavior concerning its movement. Fourthly
every behavior shows a certain degree of imperfection or irregularity.
All these reasons lead to deviations from the optimal behavior 
and may be handled as fluctuations.
\par
Nevertheless, the model gives a good impression of pedestrian movement:
Firstly there is a tendency of pedestrians to move with an intended
velocity (i.e. with an intended speed into an intended direction) (sect. 2.1).
Secondly individuals sometimes like to approach or avoid certain
objects or persons, which can be interpreted as
attractive or repulsive effects (sect.
2.2). Especially, there is a necessity of 
avoiding the collision with obstacles and other pedestrians (sect.
2.2,(b)). The consequences of each aspect will be discussed in section 3
and can be compared directly with empirical observations. Some of them
will be demonstrated by computer simulations (sect. 4).

\section{The model}

\subsection{Intended velocity of motion} \label{Motion}

\begin{enumerate}
\item[(a)] If an individual $i$ wants to arrive at a {\em destination} 
$\vec{x}_i^0$ at time $T_i$, being at time $t$ at place 
$\vec{x}_i(t)$, its {\em ideal velocity} $\vec{u}_i^0(t)$ of movement will
normally have the following properties (assuming a rectilinear way to the
destination as easiest situation first):
\begin{itemize}
\item For convenience (in order to avoid 
deceleration and acceleration processes),
the speed should be as uniform as possible, i.e. 
\begin{displaymath}
u_i^0(t) \approx const. 
\end{displaymath}
\item In walking the remaining distance
\begin{displaymath}
s_i(t) := \| \vec{x}_i^0 - \vec{x}_i(t) \|
\end{displaymath}
one should just use the remaining time $T_i - t$
(if one wants to avoid coming too late or too soon), i.e. 
\begin{displaymath}
u_i^0(t) := \frac{s_i(t)}{T_i - t} \, .
\end{displaymath}
\item The direction $\vec{e}_i$ of moving should in the simplest case
be {\em directly} oriented towards the
destination $\vec{x}_i^0$, i.e.
\begin{displaymath}
\vec{e}_i := \frac{\vec{x}_i^0 - \vec{x}_i(t)}{\|\vec{x}_i^0 -
\vec{x}_i(t)\|} \, .
\end{displaymath}
\end{itemize}
All these properties are fulfilled by the ideal velocity
\begin{equation}
\vec{u}_i^0(t) = \frac{\vec{x}_i^0-\vec{x}_i(t)}{T_i-t}
= \frac{s_i(t)}{T_i - t} \vec{e}_i \, . \label{intended}
\end{equation}
Intending to move with velocity $\vec{u}_i^0(t)$
guarantees a uniform movement and, when suffering deviations
or delays, an orientation towards the destination and an adaptation of
speed. If the available
way to the destination is not rectilinear, it can be
approximated by a polygon with edges $\vec{x}_i^n$, \dots, $\vec{x}_i^0$,
where $\vec{x}_i^n$ denotes the starting point. 
In that case, the formulas above remain unaltered, but the direction
$\vec{e}_i := \vec{e}_i^j$ of movement is oriented towards the next edge
$\vec{x}_i^{j}$, after having passed the edges
$\vec{x}_i^n$, \dots, $\vec{x}_i^{j+1}$:
\begin{displaymath}
\vec{e}_i^j := \frac{\vec{x}_i^j - \vec{x}_i(t)}{\|\vec{x}_i^j -
\vec{x}_i(t)\|} \, . 
\end{displaymath}
Now we assume that an individual $i$ of mass $m_i$, 
if moving with velocity $\vec{v}_i(t):= d\vec{x}_i(t)/dt$, applies a 
force 
\begin{equation}
 \vec{f}_i(t) \equiv m_i \frac{d\vec{v}_i(t)}{dt}
:= \gamma_i [\vec{v}_i^0(t)-\vec{v}_i(t)] 
 \label{force}
\end{equation}
to get the acceleration $d\vec{v}_i(t)/dt$ towards the
{\em intended velocity} of motion
\begin{equation}
 \vec{v}_i^0(t) := \vec{e}_i \cdot \left\{
\begin{array}{lll}
u_i^{min} & \mbox{for} & u_i^0(t) < u_i^{min} \\
u_i^0(t) & \mbox{for} & u_i^{min} \le u_i^0(t) \le u_i^{max} \\
u_i^{max} & \mbox{for} & u_i^0(t) > u_i^{max} \, .
\end{array} \right. \label{cutoff}
\end{equation}
According to this assumption, the force $\vec{f}_i$ 
is proportional to the discrepancy $\vec{v}_i^0
-\vec{v}_i$ between intended and actual velocity, and it vanishes, when 
both are equal ($\vec{v}_i=\vec{v}_i^0$). By (\ref{force}) 
$\gamma_i \vec{v}_i(t)$ approaches $\gamma_i \vec{v}_i^0(t)$
exponentially with a relaxation time of $m_i/\gamma_i$. The quantity
$\gamma_i \vec{v}_i^0$ has the
meaning of the {\em motivation to get ahead} with velocity $\vec{v}_i^0$.
For $\vec{v}_i^0$ we have introduced
a cutoff at $u_i^{max}$ and $u_i^{min}$, because velocities above 
$u_i^{max}$ are felt strenuous or uncomfortable, and velocities less than
$u_i^{min}$ are felt ``boring''. $u_i^{min}$ depends on the
surroundings (see (d)). In the following we will 
assume the common case $\vec{v}_i^0
= \vec{u}_i^0$ (i.e. $u_i^{min} \le u_i^0 \le u_i^{max}$), if nothing
contrary is mentioned.
\par
There are some other types of pedestrian movement which can be
formally reduced to type (a):

\item[(b)] Suppose that individual $i$ has the {\em plan} to pass at times
$t$ through certain places
$\vec{x}_i^0(t)$. Its intended velocity would then be
\begin{displaymath}
\vec{v}_i^0(t) = \frac{d\vec{x}_i^0(t)}{dt} \, .
\end{displaymath}
But if the individual has, due to delays, at a certain time $t_i$
still a distance $\Delta s_i(t_i)=\| \vec{x}_i^0(t_i)
- \vec{x}_i(t_i) \|$ from its intended place
$\vec{x}_i^0(t_i)$, it will try to 
make up for this distance during a time interval $\Delta t_i$,
i.e. until time $t_i+ \Delta t_i$. In that case, the 
intended velocity will, according to (\ref{intended}), 
be modified to
\begin{eqnarray}
 \vec{v}_i^0(t) &=& \frac{\vec{x}_i^0(t_i+\Delta t_i) 
-\vec{x}_i(t)}{(t_i+\Delta t_i)-t} \nonumber \\
&=& \frac{\vec{x}_i^0(t_i+\Delta t_i) 
-\vec{x}_i^0(t)}{(t_i + \Delta t_i) - t}
 + \frac{\vec{x}_i^0(t) 
-\vec{x}_i(t)}{(t_i + \Delta t_i) - t} \nonumber \\
&\approx & \frac{d\vec{x}_i^0(t)}{dt} + \frac{\vec{x}_i^0(t) 
-\vec{x}_i(t)}{(t_i + \Delta t_i) - t} \, . \nonumber 
\end{eqnarray}

\item[(c)] If an individual $i$ intends to move with {\em constant
velocity} $v_i^0$, we get type (1) by the identification 
\begin{displaymath}
 u_i^{max} := v_i^0 \, .
\end{displaymath}

\item[(d)] Suppose individual $i$ moves at leisure. Then it moves with 
a velocity
\begin{displaymath}
 v_i^0(t) = u_i^{min} (\vec{x}_i(t)) \, ,
\end{displaymath}
allowing to make as many interesting perceptions per time unit as intended.
Therefore the appropriate velocity will depend 
on the actual place $\vec{x}_i(t)$. 
The intended direction $\vec{e}_i(t)$ of movement
is given by spontaneous decisions (see section \ref{decisions}).
\end{enumerate}

\subsection{Contradictory motivations and decisions}\label{decisions}

An object or individual $j$ sometimes induces 
a psychic reaction in a pedestrian $i$, 
motivating $i$ to approach or avoid $j$ \cite{Behav}.
These attractive or repulsive effects can be described by quantities
$\vec{f}_{ij}^a$ or $\vec{f}_{ij}^r$ respectively, known as gradient of
approach or avoidance. $\vec{f}_{ij}^{a/r}$
are not forces yet, but they are a measure for the direction and strength
of the psychic motivation of $i$ to approach or avoid $j$. The strength
$f_{ij}^{a/r}$ of these motivations will lessen with increasing distance
$r_{ij} = \| \vec{x}_j - \vec{x}_i \|$ of $i$ and $j$, whereas the direction
$\vec{e}_{ij}$ will be normally oriented towards or away from $j$ , i.e.
\begin{displaymath}
\vec{e}_{ij} = \pm \widehat{\vec{r}_{ij}} = \pm \frac{\vec{r}_{ij}}{r_{ij}}
:=\pm  \frac{\vec{x}_j - \vec{x}_i}{\|\vec{x}_j - \vec{x}_i\|} 
\end{displaymath}
(+: attractive case, --: repulsive case). So with 
\begin{displaymath}
\vec{r}_{ij}:= \vec{x}_j - \vec{x}_i
= r_{ij} \cdot \widehat{\vec{r}_{ij}} 
\end{displaymath}
we find
\begin{equation}
 \vec{f}_{ij}^{a/r}(\vec{r}_{ij}) = \pm f_{ij}^{a/r}(\vec{r}_{ij}) \cdot
\widehat{\vec{r}_{ij}} \, . \label{for}
\end{equation}
In the absence of other motivations, the total effect
\begin{displaymath}
\vec{f}_{ij}(\vec{r}_{ij}) := \vec{f}_{ij}^a(\vec{r}_{ij})
+ \vec{f}_{ij}^r(\vec{r}_{ij})
\end{displaymath}
induced by $j$ would 
play an analogous role as the motivation $\gamma_i \vec{v}_i^0$ to get
ahead in equation (\ref{force}) \cite{Miller1,Miller2}.
This would lead to a movement according to
\begin{equation}
m_i \frac{d\vec{v}_i(t)}{dt} = \vec{f}_i(t)
:= \vec{f}_{ij}(\vec{r}_{ij}(t)) - \gamma_i \vec{v}_i(t) \, .
\label{att}
\end{equation}
If individual $i$ is subject to {\em a couple} of motivations, the total effect
would be the sum of all, resulting in the following equation of motion
generalizing (\ref{force}) and (\ref{att}):
\begin{equation}
m_i \frac{d\vec{v}_i(t)}{dt} =
\vec{f}_i(t) :=\left( \sum_j \vec{f}_{ij}(t) + \gamma_i \vec{v}_i^0(t) \right)
- \gamma_i \vec{v}_i(t) \, .
\label{sum}
\end{equation}
But often it is not optimal to behave according to 
(\ref{sum}), namely in the case of contradictory
motivations $\vec{f}_{ij}$, 
$\gamma_i \vec{v}_i^0$, which evoke
a psychic {\em conflict}. Then it will be better for the individual to
take a {\em decision},
whereby the behavioral alternative with the maximal
{\em utility} will be prefered
\cite{utility1,utility2}. In some cases this behavioral alternative can
be a {\em compromise}. In other cases, 
namely when the alternatives in question mutually exclude each
other, it will correspond to the alternative
which provides the {\em strongest} motivation.
We now follow {\sc Lewin}s 
``field theoretical'' view \cite{Lewin}:
Once a decision is taken, a {\em new} motivation 
\begin{displaymath}
\vec{f}_i^0(t) \equiv 
\vec{f}_i^0\left(\vec{f}_{ij}(t),\gamma_i
\vec{v}_i^0(t),t\right)
\end{displaymath}
arises as a substitute of the original motivations
$\vec{f}_{ij}$, $\gamma_i\vec{v}_i^0$. This motivation is 
some kind of {\em psychic tension},
which causes the individual to act towards its aim in order to {\em diminish}
this tension. 
In the case of pedestrians, the body
will be induced to generate a {\em physical force}
\begin{displaymath}
\vec{f}_i(t) :=  \vec{f}_i^0\left(\vec{f}_{ij}(t),\gamma_i
\vec{v}_i^0(t),t\right) - \gamma_i \vec{v}_i(t) \, , 
\end{displaymath}
which then causes a
movement according to
\begin{equation}
 m_i \frac{d\vec{v}_i(t)}{dt} = \vec{f}_i(t) 
= \vec{f}_i^0\left(\vec{f}_{ij}(t),\gamma_i
\vec{v}_i^0(t),t\right)
 - \gamma_i \vec{v}_i(t) \label{new}
\end{equation}
(compare to (\ref{force}), (\ref{sum})). 
Due to (\ref{new}),
a pedestrian will stop moving only, when the motivation to move is
vanishing, i.e. when
\begin{equation}
 \vec{f}_i^0\left(\vec{f}_{ij}(t),\gamma_i
\vec{v}_i^0(t),t\right)  = \vec{0} \, . \label{eq1}
\end{equation}
By 
\begin{displaymath}
\vec{f}_i^0\left(\vec{f}_{ij}(t),\gamma_i
\vec{v}_i^0(t),t\right)
:= \sum_j \vec{f}_{ij} + \gamma_i\vec{v}_i^0
\end{displaymath}
(\ref{sum}) can be interpreted as special case of (\ref{new}), being valid 
as long as no decision is taken.
In that case (\ref{eq1}) has the form of an equilibrium condition for the
motivations
$\vec{f}_{ij}$, $\gamma_i\vec{v}_i^0$:
\begin{equation}
 \sum_j \vec{f}_{ij}(t) + \gamma_i \vec{v}_i^0(t) = \vec{0} \, . \label{eq}
\end{equation}
\par
Now two examples for situations will be
given, in which conflicts between several motivations occur:
\begin{enumerate}

\item[(a)] {\bf Joining behavior} \\
Suppose individual $i$ perceives an attractive object or individual
$j$ of attraction $f_{ij}^a(t_{ij})$ at time $t_{ij}$. Individual $i$
will then spontaneously decide to meet $j$, if there is enough time 
to do so. We assume this to be the case if
\begin{displaymath}
 f_{ij}^a(t_{ij}) > \gamma_i v_i^0(t_{ij}) = \gamma_i \frac{s_i
(t_{ij})}{T_i - t_{ij}} \, ,
\end{displaymath}
i.e. if the motivation $f_{ij}^a$
for joining $j$ is greater than the motivation $\gamma_i v_i^0$ to
continue walking (see (\ref{force})). 
(Here, we have made the simplification that there is 
only a small detour necessary to
meet $j$.) 
\par
Individual $i$ will stay at the meeting point 
for a time $\tau_{ij}$ and will leave at the moment $t_{ij}+\tau_{ij}$,
when the tendency $f_{ij}^a$ to join the attractive person or object $j$
becomes less than the increasing tendency $\gamma_i v_i^0$
to get ahead.
($v_i^0(t)$ is growing according to the delay 
$\tau_{ij}$ resulting from the
stay.) This condition can be written in the form
\begin{equation}
 f_{ij}^a(t_{ij}+\tau_{ij}) \stackrel{!}{=} \gamma_i v_i^0(t_{ij}+\tau_{ij})
= \gamma_i \frac{s_i(t_{ij})}{T_i - (t_{ij} + \tau_{ij})} \label{beding}
\end{equation}
(see (\ref{intended})) because of $s_i(t_{ij} + \tau_{ij}) = s_i(t_{ij})$.
By (\ref{beding}) the staying time $\tau_{ij}$ can be calculated as
\begin{equation}
 \tau_{ij} = (T_i - t_{ij}) - \frac{\gamma_i s_i(t_{ij})}{f_{ij}^a}
= (T_i - t_{ij}) \frac{f_{ij}^a - \gamma_i v_i^0(t_{ij})}{f_{ij}^a} 
\, , \label{tau}
\end{equation}
if $f_{ij}^a$ is constant with time ($f_{ij}^a(t) = f_{ij}^a$).
\par
If $f_{ij}^a(t_{ij}) \le \gamma_i v_i^0(t_{ij})$ or, 
equivalently, $\tau_{ij} \le 0$,
there is not enough time for joining $j$, and individual $i$ will do best
to continue walking without changing its way.
\par
Summarizing (a), the decision of individual $i$ leads to a new motivation
\begin{displaymath}
 \vec{f}_i^0\left( \vec{f}_{ij}^a(t), \gamma_i \vec{v}_i^0(t)
\right) := \gamma_i \vec{v}_i^0(t) \,
 \Theta\left(f_{ij}^a(t) < \gamma_i v_i^0(t)\right) \, ,
\end{displaymath}
which substitutes the contradictory motivations $\vec{f}_{ij}^a$ and
$\gamma_i \vec{v}_i^0$. Here, we have introduced the decision function
\begin{displaymath}
\Theta(x) := \left\{
\begin{array}{ll}
1 & \mbox{if $x$ is true} \\
0 & \mbox{if $x$ is false.} 
\end{array} \right.
\end{displaymath}
Of course, individual $i$ will change its direction of motion temporarily
from $\vec{e}_i(t_{ij})$ to $\widehat{\vec{r}_{ij}}$, if this is necessary
for joining $j$.

\item[(b)] {\bf Avoidance behavior} \\
Suppose individual $i$, e.g. in order to avoid a collision, 
decides at time $t_i$
to avoid an object or individual $j$ (i.e. to keep a certain distance).
Then, on one hand, individual $i$ tries to minimize the 
maximal repulsive effect against $j$, namely
\begin{displaymath}
 \max_t f_{ij}^r(\vec{r}_{ij}(t)) =: 
f_{ij}^r(\vec{r}_{ij}(t_{ij}))\, ,
\end{displaymath}
which normally occurs at the moment $t_{ij}$ of greatest approach
$r_{ij}(t_{ij})$.
On the other hand, it wants to minimize the increase of the pressure
$\gamma_i v_i^0$ to get ahead, i.e. to minimize the detour, which is
necessary to avoid $j$. The best compromise will be to take a way, for which
the maximal repulsive tendency 
and the tendency 
to get ahead have equal
amounts, namely for which
\begin{equation}
 f_{ij}^r(\vec{r}_{ij}(t_{ij})) = \gamma_i v_i^0(t_{ij}) \, , \label{glgew}
\end{equation}
and to take a rectilinear path. 
This path is given as tangent to the area
\begin{equation}
 {\cal T}_{ij}(t) := \{\vec{x}: f_{ij}^r(\vec{x}_j(t) - \vec{x}) > \gamma_i
v_i^0(t) \} \, , \label{terr}
\end{equation}
which describes the {\em territory} of $j$, that is 
{\em respected}, i.e. not entered by
individual $i$. 
Due to (\ref{terr}) 
the area of the respected territory ${\cal T}_{ij}(t)$ decreases with
increasing intended velocity $v_i^0(t)$ or, equivalently, with increasing 
pressure $\gamma_i v_i^0(t)$ to get ahead.
\par
For the sake of completion, 
we assume the following additional laws of pedestrian
avoidance behavior:
\begin{itemize}
\item When avoiding a pedestrian or obstacle $j$, individual $i$
will keep its intended speed $v_i^0(t_i)$, changing only its intended
direction from $\vec{e}_i(t_i)$ to
\begin{displaymath}
\frac{\vec{x}_i^0(t_{ij}) - \vec{x}_i(t_i)}
{\|\vec{x}_i^0(t_{ij}) - \vec{x}_i(t_i)\|} \, ,
\end{displaymath}
where $\vec{x}_i^0(t_{ij})$ is the intended position of $i$ for the
moment of greatest approach. 
According to this, the motivation to get ahead will be
changed from $\gamma_i \vec{v}_i^0(t_i)$ to 
\begin{displaymath}
 \vec{f}_i^0\left(\vec{f}_{ij}^r(t),\gamma_i\vec{v}_i^0(t)
\right) := \gamma_i
v_i^0(t) \frac{\vec{x}_i^0(t_{ij}) - \vec{x}_i(t_i)}
{\|\vec{x}_i^0(t_{ij}) - \vec{x}_i(t_i)\|} 
\end{displaymath}
during the time it takes to avoid $j$ (i.e. for times $t$ with $t_i \le t \le
t_{ij}$).

\item An individual $i$ reacts a time $\Delta t_{ij} := t_{ij} - t_i$
before a collision would be expected. This time $\Delta t_{ij}$
is a psychic parameter, which, of course,
will be the greater the larger the dimension of 
the obstacle $j$ is. The {\em distance}
$d_{ij}$ of reaction before the location of
a probable collision is 
the product of $\Delta t_{ij}$ and speed $v_i$:
\begin{displaymath}
 d_{ij} = v_i \Delta t_{ij} \, .
\end{displaymath}
It is plausible, that the necessary
angular change of direction when avoiding an obstacle $j$ will be the
greater, the lower the distance $d_{ij}$ of the obstacle $j$
is. So the (average)
change of direction will be the greater, the lower the (average) speed is.
This can be observed when comparing more and less crowded situations.

\item If the distance for passing $j$ on the left is
nearly the same as for passing $j$ on the right, we assume individual $i$
to take the right hand side with probability $p_1$ and the left hand side with
probability $p_2:=1-p_1$. 

\item But if there is no chance of passing $j$, e.g. when the
way is too crowded,
individual $i$ will decelerate (as long as necessary) 
to a velocity $\vec{v}_i$, which allows a maximal component $\vec{v}_i \cdot
\vec{e}_i$ of movement into the intended direction $\vec{e}_i$. This maximal
component is normally equal to the component $\vec{v}_j \cdot \vec{e}_i$,
which the hindering pedestrian's velocity $\vec{v}_j$ has in direction
$\vec{e}_i$ (corresponding to the situation, that individual $i$
walks in a gap behind a pedestrian $j$ with velocity $\vec{v}_j$).
However, if pedestrian $j$ has an opposite direction with respect to $i$
($\vec{v}_j \cdot \vec{e}_i < 0$), it will be better for individual
$i$ to stop ($\vec{v}_i = \vec{0}$). Summarizing these results we have the
relation
\begin{displaymath}
 \vec{v}_i\cdot\vec{e}_i :=
\left\{
\begin{array}{ll}
\vec{v}_j\cdot \vec{e}_i & \mbox{if } \vec{v}_j\cdot\vec{e}_i >0 \\
0 & \mbox{else.}
\end{array} \right.
\end{displaymath}

\end{itemize}

\end{enumerate}

\section{Conclusions and comparison with real situations}

\subsection{Effects of the intended velocity of motion}

\begin{enumerate}
\item[(a)] {\bf Velocity of motion} \\
According to (\ref{force}) a pedestrian would normally
walk with velocity $\vec{v}_i(t) \approx \vec{v}_i^0(\vec{x}_i(t))$. But
in order to avoid collisions, an individual $i$ suffers detours or delays, and
as a consequence, its {\em smoothed} velocity $\vec{\bar{v}}_i(t)$
of motion will probably have the more general form
\begin{equation}
\frac{d\vec{\bar{x}}_i(t)}{dt}=
 \vec{\bar{v}}_i(t) \approx \vec{w}_i + k_i \vec{v}_i^0(\vec{\bar{x}}_i(t))
= \vec{w}_i + k_i \frac{\vec{x}_i^0
- \vec{\bar{x}}_i(t)}{T_i - t} \label{motion}
\end{equation}
with $k_i \le 1$ (see (\ref{intended})). 
$k_i$ and $\vec{w}_i$  are empiric parameters depending on the walking
situation and describing the effect of ``interindividual interactions''.
(\ref{motion}) can be solved by
\begin{displaymath}
\vec{\bar{v}}_i(t) = \vec{w}_i + \left\{
\begin{array}{lll}
\left(\frac{k_i}{T_i-t_i^0}[\vec{x}_i^0-\vec{\bar{x}}_i(t_i^0)]
+ \frac{k_i}{k_i-1} \vec{w}_i\right)
\left(1-\frac{t-t_i^0}{T_i-t_i^0}\right)^{k_i-1} - \frac{k_i}{k_i-1}\vec{w}_i
& \mbox{if} & k_i \ne 1 \\
\frac{1}{T_i-t_i^0}[\vec{x}_i^0 - \vec{\bar{x}}_i(t_i^0)]
+ \ln \left( 1 - \frac{t-t_i^0}{T_i - t_i^0} \right)\cdot \vec{w}_i
& \mbox{if} & k_i = 1\, ,
\end{array} \right.
\end{displaymath}
where $t_i^0$ is the time when individual $i$ starts walking.
We can conclude the following:
\begin{itemize}
\item If the smoothed actual velocity $\bar{v}_i$ is less than the intended
velocity $v_i^0$, then $\bar{v}_i$ and $v_i^0$ will be growing 
with time, because from (\ref{motion}) 
\begin{displaymath}
\frac{d\vec{v}_i(t)}{dt} \approx k_i \frac{\vec{v}_i^0(t) - \vec{v}_i(t)}
{T_i - t}  = k_i \frac{d\vec{v}_i^0(t)}{dt}
\end{displaymath}
can be derived.
So individual $i$ will speed up in the course of time
unless the maximal velocity $\vec{v}_i^{max} =
\vec{w}_i + k_i u_i^{max} \vec{e}_i$ is reached. (Apart from (\ref{motion})
we have now taken into account eq. (\ref{cutoff}).)

\item Individual $i$ will arrive at destination $\vec{x}_i^0$ too late if
the smoothed actual velocity $\bar{v}_i(t)$ would have to exceed the maximal
velocity $v_i^{max}$ before time $T_i$, i.e. if
\begin{displaymath}
 \lim_{t\rightarrow T_i}\bar{v}_i(t) > v_i^{max} \, .
\end{displaymath}
 \item It will keep less distance to other pedestrians $j$ as $\bar{v}_i(t)$
increases (see sect. \ref{decisions},(b)), 
because of
\begin{displaymath}
 \gamma_i \vec{v}_i^0(t) = \gamma_i \frac{\vec{\bar{v}}_i(t) - \vec{w}_i}
 {k_i} 
\end{displaymath}
and equation (\ref{glgew}).
Individual $i$ then shows less
respect against the ``territory'' of an individual $j$: it walks more
aggressively and perhaps even pushes.

\item In crowded situations individual $i$ can prevent having to hurry by
intending to walk with velocity 
\begin{displaymath}
 \vec{v}_i^{0} := - \frac{\vec{w}_i}{k_i} + \frac{1}{k_i}
\frac{\vec{x}_i^0 - \vec{\bar{x}}_i(t)}{T_i - t} \, .
\end{displaymath}
This strategy will lead to a smoothed actual velocity of 
$\vec{\bar{v}}_i = \vec{v}_i^0$.
\end{itemize}

\item[(b)] {\bf Effect of an unexpected detour} \\
In some situations an individual $i$ has to walk an unexpected 
detour $\Delta s_i$,
e.g. if it has forgotten something and is suddenly remembering this at
time $t_i^+$. So the intended velocity changes according to (\ref{intended}) 
from 
\begin{displaymath}
 v_i^0(t_i^-) = \frac{s_i(t_i^-)}{T_i-t_i^-}
\end{displaymath}
at the preceding moment $t_i^-$ to 
\begin{displaymath}
 v_i^0(t_i^+) = \frac{s_i(t_i^+)}{T_i - t_i^+} 
= \frac{s_i(t_i^-) + \Delta s_i}{T_i - t_i^+} > v_i^0(t_i^-) \, .
\end{displaymath}
By (\ref{force}) this gives rise to a
sudden increase of velocity $v_i$, which can often be observed, especially
for individuals who walk according to a plan $\vec{x}_i^0(t)$
(see sect. \ref{Motion},(b)). These individuals try to speed up to maximal
velocity $v_i^0(t_i^+):=u_i^{max}$ 
until they have, after a time interval
\begin{displaymath}
 \Delta t_i \ge \frac{\Delta s_i}{u_i^{max} - v_i^0(t_i^-)} \, ,
\end{displaymath}
reached their plan $\vec{x}_i^0
(t_i + \Delta t_i)$ again 
(in the sense of $\vec{x}_i(t_i + \Delta t_i)
= \vec{x}_i^0(t_i + \Delta t_i)$).

\item[(c)] {\bf Behavior in a queue} \\
If the front of a queue has come to rest, the following phenomenon
can often be observed: After a while, one of the waiting individuals
begins to move forward a little, causing the successors
to do the same. This process propagates in a
wave-like manner to the end of the queue,
and the distance to move forward increases.
\par
Why do individuals behave in such a paradox 
way?---They don't get away any faster
but only cause the queue to become more crowded! Our model gives the
following interpretation:
\par
At time $t_i$ an individual $i$ keeps a distance $r_{i,i-1}(t_i)$
to the individual $i-1$ in front, which is 
(according to (\ref{eq}) and (\ref{glgew}))
given by
\begin{displaymath}
 f_{i,i-1}^r(r_{i,i-1}(t_i)) = \gamma_i v_i^0(t_i) \, .
\end{displaymath}
$f_{i,i-1}^r$ is the repulsive effect describing the territory
of individual $i-1$ respected by $i$. 
As we know from (\ref{intended}), $v_i^0(t)$ grows
as time $t$ passes, because individual $i$ is at rest ($\vec{x}_i(t)
= \vec{x}_i(t_i)$). So at time $t_i + \Delta t_i$ individual $i$ would prefer
to have a distance
\begin{displaymath}
 r_{i,i-1}(t_i+\Delta t_i) =: r_{i,i-1}(t_i) - \Delta r_i
(t_i + \Delta t_i) \, ,
\end{displaymath}
which has reduced by an amount $\Delta r_i$ and is given by
\begin{equation}
 f_{i,i-1}^r(r_{i,i-1}(t_i+\Delta t_i)) = \gamma_i v_i^0(t_i + \Delta t_i) \, .
\label{move}
\end{equation}
But individual $i$ moves up a distance $\Delta r_i$ only if
\begin{equation}
 \Delta r_i \ge \Delta r_i^{min}\, , \label{incr}
\end{equation}
i.e. if the increment $\Delta r_i$ exceeds a minimal 
stride $\Delta r_i^{min}$. So the first individual
moving up is the individual $i$, for which condition
\begin{displaymath}
 \Delta r_i(t_i + \Delta t_i) = \Delta r_i^{min}
\end{displaymath}
is fulfilled first. This is the case
at a time $t := t_i + \Delta t_i$, i.e. a time
interval $\Delta t_i$ after its last step at time $t_i$. Now the successors
$i+n$ $(n\ge 1)$ will move forward a distance
\begin{displaymath}
 s_{i+n} = \sum_{j=i}^{i+n} \Delta r_j(t) 
= \sum_{j=i}^{i+n} \Delta r_j(t_j + \Delta t_j)
\end{displaymath}
according to (\ref{move}) and (\ref{incr}), 
because $s_{i+n} \ge \Delta r_{i+n}^{min}$
will normally be fulfilled.
\end{enumerate}

\subsection{Attractive and repulsive effects}

\begin{enumerate}
\item[(a)] {\bf Constant density} \\
Suppose a number of $N$ individuals having only a negligible intention to
move ($\vec{v}_i^0 \approx \vec{0}$) stay in an area of a (dining) hall,
a waiting room, a beach, an underground station, etc. with size $A$. 
One can observe then
a quite uniform distribution of individuals (with constant density
$N/A$) if there are no special attractions in area $A$ and no 
aquaintances between the individuals (see (b)). This is due to
the repulsive effects $\vec{f}_{ij}^r$ between 
each pair of individuals $i$ and $j$, which
are in equilibrium (see (\ref{eq})), when all individuals
occupy a personal territory of nearly equal size. 

\item[(b)] {\bf Formation of groups} \\
If there {\em are} aquaintances between the individuals of example (a),
a truncated {\sc Poisson} distribution
\begin{equation}
 p_k = {\cal N} \frac{\lambda^k}{k!} \qquad k=1,2,\dots \label{distrib}
\end{equation}
can be found for the proportion $p_k$ of groups consisting of $k$ members.
This distribution is well confirmed by empirical data \cite{Coleman2}
and can be 
explained by the following mathematical model of {\sc Coleman}
\cite{Coleman1,Coleman2}:
\begin{eqnarray}
 \frac{dp_k}{dt} &=& \mbox{[transitions from $l(\ne k)$ to $k$ } -
\mbox{ transitions from $k$ to $l(\ne k)$]/time unit} \nonumber \\
&=& \sum_{l(\ne k)} p_l \cdot r(l\rightarrow k) - \sum_{l(\ne k)}
p_k \cdot r(k \rightarrow l) \nonumber \\
&=& (p_{k+1} \cdot (k+1)\cdot \beta + p_{k-1} \cdot \alpha \cdot p_1)
- (p_k \cdot k \cdot \beta + p_k \cdot \alpha \cdot p_1) 
\label{trans}
\end{eqnarray}
for $k=2,3,\dots$, and
\begin{equation}
 \sum_{k=1}^\infty p_k = 1 \, . \label{norm}
\end{equation}
In (\ref{trans}) we 
have used
\begin{displaymath}
 r(k\rightarrow l) = \left\{
\begin{array}{ll}
k \cdot \beta & \mbox{if } l=k-1 \\
\alpha \cdot p_1 & \mbox{if } l=k+1 \\
0 & \mbox{else}
\end{array} \right.
\end{displaymath}
with $l \ge 2$.
This means that a group with $k$ individuals loses individuals
independently with rate $\beta$ and gains single individuals with 
rate $\alpha\cdot p_1$ (which is proportional to the number of single
individuals). Other transitions are assumed to be relatively
unimportant.
\par
(\ref{trans}), (\ref{norm}) have the stationary solution
\begin{displaymath}
 p_k = \frac{1}{\mbox{e}^\lambda - 1} \frac{\lambda^k}{k!} \, ,
\end{displaymath}
given by $dp_k/dt = 0$, where
\begin{equation}
 \lambda := \ln \left( \frac{\alpha}{\beta} + 1 \right) \, .\label{lambda}
\end{equation}
We now connect these results with our model:
For $\beta$ we could simply take the mean value of the reciprocal 
$1/\tau_{ij}$ of the
time $\tau_{ij}$ which an individual $i$ stays in a group $j$, because
this is the rate of leaving a group
(see (\ref{tau})):
\begin{equation}
 \beta := E\left(\frac{1}{\tau_{ij}}\right) \, .
\end{equation}
On the other hand, $\alpha$ can be assumed of the form
\begin{equation}
 \alpha := p_+ J \, , \label{alpha} 
\end{equation}
where $J$ is the rate of recognized groups per time unit and
$p_+$ is the probability to {\em join} a recognized group $j$. 
According to sect. \ref{decisions},(a), 
$p_+$ is the probability $P(\tau_{ij} >0)$, that the staying time
$\tau_{ij}$ is positive:
\begin{equation}
 p_+ := P(\tau_{ij} > 0 ) = P(f_{ij}^a > \gamma_i v_i^0) \, . \label{pplus}
\end{equation}
$f_{ij}^a$ is, of course, 
the attractive effect between individual $i$ and group $j$.
\par
Due to (\ref{lambda}) to (\ref{pplus}) the following conclusions 
can now be made:
\begin{itemize}
\item Parameter $\lambda$, which is a measure for the average number of
members of a group, increases with the mean value of the staying time
$\tau_{ij}$, i.e. it decreases with growing intended velocity $v_i^0$
and increases with growing remaining time $T_i - t_{ij}$ (see (\ref{tau})). 
This is consistent with the data \cite{Coleman2}.

\item If the motivation $f_{ij}^a$ to join a group $j$ is less than
the motivation $\gamma_i v_i^0$ to get ahead for all individuals $i$ and 
groups $j$,
we have $p_+ = 0$ and $\alpha = 0$. In that case no groups are forming at all
and (a) can be applied again (if $\vec{v}_i^0 \approx \vec{0}$). 
\end{itemize}

\item[(c)] {\bf 
Superposition of attractive and repulsive effects} \\
Often a person or object $j$ has an attractive effect $\vec{f}_{ij}^a$
and a repulsive effect $\vec{f}_{ij}^r$ as well. As a consequence of 
equation (\ref{att}), individual $i$ will then show one of several
characteristic dynamic behaviors known from approach-avoidance conflicts,
depending on the special form of the motivation gradient $\vec{f}_{ij}
(\vec{r}_{ij})$ \cite{Herk}. Especially, for negligible intention to 
move ($\vec{v}_i^0 \approx \vec{0}$), individual $i$ will prefer a certain
distance \cite{Miller1,Party}, for which the equilibrium condition
\begin{displaymath}
f_{ij}(\vec{r}_{ij}) = 
f_{ij}^a(\vec{r}_{ij}) - f_{ij}^r(\vec{r}_{ij}) = 0
\end{displaymath}
is fulfilled (see (\ref{eq}) and (\ref{for})), i.e. for which the
attractive and the repulsive effect have equal strengths.

\item[(d)] {\bf Break of symmetry for avoidance behavior} \\
Suppose two individuals walk in opposite direction and try to avoid
each other in order not to suffer a collision. 
Then each tries to pass the other with probability $p_1$ on the
right and probability $p_2=1-p_1$ on the left (see sect. \ref{decisions},(b)).
\par
The probability for avoiding each other successfully is then
\begin{displaymath}
 p_1\cdot p_1 + p_2 \cdot p_2 =: 1 - w \, .
\end{displaymath}
Otherwise, with probability
\begin{equation}
 w = p_1 \cdot p_2 + p_2 \cdot p_1 = 2p_1 \cdot p_2 \le \frac{1}{2} \, ,
\label{we}
\end{equation}
they have to try again, etc., until they pass on {\em different} sides. 
This phenomenon is well known. 
\par
The mean value $E(n)$ for the necessary number $n$
of attempts to avoid each other
is given by
\begin{equation}
 E(n) = \sum_{n=1}^\infty n \cdot w^{n-1}\cdot (1-w) = \frac{1}{1-w} \, .
\label{En}
\end{equation}
Taking (\ref{we}) into account,
this expression is {\em maximal} for $w=1/2$, i.e.
for symmetric probabilities 
\begin{displaymath}
 p_1 = p_2 = \frac{1}{2}
\end{displaymath}
of avoidance for both sides. (\ref{En}) is {\em minimal} 
for $p_1=0$ or $p_1=1$ (deterministic behavior!).
Therefore {\em asymmetric} probabilities $p_1 \ne p_2$ of avoidance are
favourable. In fact, in most countries individuals more frequently
pass other individuals on the right ($p_1 > 1/2$). As a consequence, crowded
ways often show two different lanes of opposite direction, which
stick to the right side respectively \cite{Soldaten,bbi,Geschwindigkeit}. 
This behavior reduces the frequency of  
situations of avoidance and corresponding delays.
\par
{\bf Selection of one behavioral alternative} \\
For explanation of the break of symmetry ($p_1 \ne p_2$), we consider the
following general model which describes the temporal change
of the proportion $p_k$ of individuals showing a certain 
behavioral alternative
$k$ (compare to \cite{Eigen}):
\begin{eqnarray}
 \frac{dp_k}{dt} &=& \sum_{l(\ne k)} (M_{kl}p_l - M_{lk}p_k) \nonumber \\
                 &+& s_k p_k + \xi_k \, .
\label{behav}
\end{eqnarray}
$M_{kl}$ are the {\em mutation} rates for changes
from behavior $l$ to behavior $k$ per
time unit and person. For the choice
\begin{equation}
 s_k := M_{kk} - \sum_l M_{ll} p_l \, ,
\label{select}
\end{equation}
$s_kp_k$ has the effect of a {\em selection} 
between the behavioral alternatives
$k$. $\xi_k$ are random fluctuations of the proportion $p_k$.
%
%
\par
For the problem of avoidance we have 
only two alternatives: one to pass a hindering
pedestrian on the
right ($k:=1$), 
and the other to pass it on the left ($k:=2$). As mutation matrix we take
\begin{equation}
 \underline{M} := \underline{A} + \underline{B}
\end{equation}
with 
\begin{equation}
 \underline{A} := \lambda \left(
\begin{array}{cc}
p_1 & 1-p_2 \\
1-p_1 & p_2
\end{array} \right) 
\end{equation}
and
\begin{equation}
 \underline{B} := \beta \left(
\begin{array}{cc}
1/2 & 1/2 \\
1/2 & 1/2 
\end{array} \right) \, . \label{be}
\end{equation}
According to $\underline{A}$, a behavioral alternative $k$ becomes more
probable (by learning), the greater the proportion $p_k$ of individuals
with behavior $k$ is (because in our case behavior $k$ is the more
{\em successful} the more often it occurs) \cite{Lernen1,Lernen2}. 
On the other hand, $\underline{B}$
describes a random choice of some behavior $k$ with probability $1/2$
due to trial (and error). (The individual behavior depends on the respective
situation.)
\par
Substitution of (\ref{select}) to (\ref{be}) in (\ref{behav}) now gives
\begin{equation}
 \frac{dp_k}{dt} = [ 2 \lambda p_k\cdot(1-p_k) - \beta]\cdot
\left(p_k - \frac{1}{2} \right) + \xi_k \, , \label{change}
\end{equation}
which, for $\beta \ge \lambda/2$, has the only stationary solution $p_k = 1/2$.
However, for a low tendency $\beta$ to choose the behavior randomly
($0 \le \beta < \lambda/2$), (\ref{change}) has three stationary
solutions: $p_k = 1/2$, being {\em unstable}
against fluctuations $\xi_k$, and
$p_k = 1/2 \cdot (1 \pm \sqrt{1 - 2\beta/\lambda})$, being {\em stable}! As a
consequence of the instability of
$p_k = 1/2$, fluctuations will cause the proportion
$p_k$ to tend either towards
$p_{k} = 1/2 + 1/2 \sqrt{1-2\beta/\lambda}$ 
(prefering the right side) or towards
$p_{k} = 1/2 - 1/2 \sqrt{1-2\beta/\lambda}$ (prefering the left side). By 
spatial diffusion of this learning process the prefered behavior is 
spread over wide areas (e.g. countries) and stabilized against crossing
$p_{k} = 1/2$, which could in principal be induced by fluctuations.
\par
We now assume that an individual $i$ {\em overtakes} a pedestrian $j$ walking
in the {\em same} direction. Here, we normally do not have to expect
any complications by the behavior of $j$. So the avoidance behavior will be
successful with probability $w=1$, regardless of the side of passing.
Our mutation matrix $\underline{M}$ then will not depend on the proportions
$p_1$, $p_2$ of pedestrians passing on the left or on the 
right ($\lambda = 0$).
This time we have the equation
\begin{displaymath}
 \frac{dp_k}{dt} = - \beta \left(p_k - \frac{1}{2}\right)
+ \xi_k 
\end{displaymath}
(see (\ref{change})),
which has only one stationary solution: the symmetric probability 
$p_k = 1/2$ of avoidance, which is stable!
\end{enumerate}

\section{Computer simulations}

In order to test the somewhat algorithmical model of section 2 (especially
section 2.2,(b)), some simple computer simulations have been carried out.
The corresponding computer program works as follows:
\begin{itemize}
\item First the geometrical configuration is determinded
(e.g. a normal pedestrian way or a pedestrian way with several obstacles).

\item In the examples presented, two types (i.e. main directions)
of motion are necessary: Pedestrians intending to walk from the left to the
right are represented by black lines, those intending to walk in the opposite
direction are represented by grey lines. Every line has the meaning of an
individual's actual stride, and its length is proportional to its velocity.

\item As initial configuration a statistically uniform spatial distribution
of $N$ pedestrians is taken ($N=350$ or 500), one half belonging to the
black type of motion, the other half belonging to the grey type
(see fig. 1). The intended speeds
of each direction are distributed by chance ({\sc Gauss}ian), whereby the same
mean speeds and the same
velocity variances were chosen for both directions of motion.

\item At the beginning of the simulation, a certain order of the $N$
pedestrians is chosen at random. The pedestrians
take each step according to that order.
After even the $N$th pedestrian has taken its $S$th step, the 1st
pedestrian is taking its $(S+1)$st one.
For each individual leaving on one side of a figure, an equivalent one
enters on the other side, i.e.
the right side of each figure can be assumed to be connected to the left side
(periodic boundary conditions).

\item Now the considerations from section 2.2,(b) are taken into account: 
A pedestrian taking its next step will move by its intended stride
into its intended direction, if this is possible. If not, i.e. if it would
have to cross another pedestrian's  step, it will change its direction by
an angle, which will be the greater, the nearer the hindering pedestrian is.
However, if even this does not prevent him from crossing
another pedestrian's step, the intended stride will be
taken as short as necessary, possibly leading to a stop.
In the case of a change of direction, the right side is chosen with probability
\begin{displaymath}
 p_1 := \left\{
\begin{array}{ll}
1/2 & \mbox{if both pedestrians belong to the same direction of motion} \\
p & \mbox{if the pedestrians belong to different directions of motion} \, .
\end{array} \right.
\end{displaymath}
The left side is chosen with probability $p_2 = 1 - p_1$.

\item If a pedestrian comes into the proximity of an obstacle, 
it temporarily changes its
intended direction. It prefers to pass the obstacle
at the nearest side in order to suffer the least possible detour. If both sides
have approximately the same distance, each side is chosen with probability
$1/2$.
\end{itemize}
The computer simulations show the following results:
\begin{itemize}
\item For symmetric avoidance behavior ($p=1/2$), changes of direction
appear very often, because encounters of pedestians from opposite directions
are likely to 
happen everywhere (see fig. 2). In the case of {\em asymmetric}
avoidance behavior ($p=0.7$), two walking lanes of opposite direction develop
in the course of time (see fig. 3). Obviously, there
are less changes of direction necessary, 
occuring mainly at the borderline between the
opposite lanes.

\item In the presence of an obstacle, a pedestrian free area develops in front
of and behind the obstacle (see figures 4 and 5). But, 
whereas an obstacle in the
middle of a pedestrian way 
causes only a small area not to be used (see fig. 4),
obstacles at the margin do reduce the effective width over a long distance
(see fig. 5).
\end{itemize}

\section{Conclusions}
We have set up a model for the movement of pedestrians starting from the
idea that individual decisions are guided by maximization of utility.
Once a decision is taken, a special kind of psychic motivation or tension to
realize this decision arises, which causes the individual to act towards
its aim in order to neutralize the psychic tension. For example, when
an individual $i$ wants to reach a certain destination at a time $T_i$, it
would do best to walk with a suitable velocity 
$\vec{v}_i^0$. So the pedestrian will
decide to walk with the ``intended velocity'' $\vec{v}_i^0$, causing it
to apply a physical force $\vec{f}_i$, which vanishes, when the pedestrian's
actual velocity $\vec{v}_i$ is equal to the intended one. 
In the case of delays,
the intended velocity
has to be corrected upwards in the 
course of time, causing the pedestrian to speed
up and perhaps to walk more aggressively. Waiting
in a queue that has come to rest, an individual will instead move forward after
some time, which is motivating the successors to move forward, too. Therefore,
this behavior propagates in a wave-like manner to the end of the queue and
leads to a more crowded queue.
\par
In addition, a pedestrian is subject to attractive or repulsive
influences, motivating it to
approach or to avoid certain individuals or things $j$.
If, for example, the motivation
$\vec{f}_{ij}^a$
to approach some person (say a friend) or some object (e.g. a shop-window)
is greater than the motivation to get ahead,
the pedestrian $i$ will decide to join this individual or object for a while.
But it will leave the moment at which the motivation to join the attractive
person or object $j$ becomes less than 
the increasing motivation to get ahead with the
intended velocity (which is growing according to the delay resulting from the
stay). If, right from the beginning, the motivation of a pedestrian 
to get ahead is greater than the motivation to join a certain person or object 
$j$, 
the pedestrian's best decision will be not to change its path at all.
This model leads to a detailed description of group formation.
\par
However, there are also repulsive effects $\vec{f}_{ij}^r$. 
They describe, for example, the
personal territories of individuals $j$. 
As a consequence, individuals who don't 
know each other normally spread uniformly
in an area of a hall, a waiting room, a cafe, a beach, etc. (if there
are no special attractions). In situations where a pedestrian $i$ has to avoid
another one $j$ in order to prevent a collision, it prefers to
suffer only a minimal detour. So individual $i$ will pass individual $j$ along
a tangent 
to the territory of $j$ respected
by $i$. This respected territory is given as the area around $j$, for which
the repulsive effect $f_{ij}^r$ of $j$ is greater than the
motivation $\gamma_i v_i^0$ of $i$ to get ahead 
with speed $v_i^0$. 
\par
Mathematically, it appears to be
favourable when most pedestrians prefer either the right side or the left side
when passing each other. This results in the development of walking
lanes in pedestrian crowds. With both sides being equivalent, 
one side will be used by a growing majority, once it has been chosen at
random.
This is one example being representative for many others, 
where the most successful
or most efficient behavior is adopted by trial and error
causing a selection between behavioral alternatives.
\par
After having set up a ``microscopic'' model, i.e. one 
for the movement of {\em individuals}, one
may be interested in a model for a great number of interacting pedestrians.
Such a model is developed in \cite{Helbing}. It shows some similarities to
gaskinetic and fluid dynamic equations, but contains some additional terms that
are characteristic for pedestrian movement.
\par
{\bf Acknowledgements:} \\
First I want to thank Prof. Dr. M. R. Schroeder to give me the chance
to work on an interdisciplinary field: modelling the {\em social}
behavior of pedestrians by {\em mathematical} models. Secondly I am
grateful to Prof. Dr. R. Kree, Prof. Dr. W. Scholl, Prof. Dr. W.
Weidlich, Dr. habil. G. Haag and Dr. R. Reiner 
for their stimulating discussions.
Last but not least I'm obliged to D. Weinmann, N. Empacher, K.-G. Haas and
E. Webster for reading, correcting
and commenting my manuscripts.

\clearpage
\thispagestyle{empty}
\vspace*{8.7cm}
{\bf Fig. 1} ($N=500$, $S=0$): \\
Initial configuration: $N$ pedestrians with varying speeds are
distributed randomly
over a pedestrian way, the black ones walking from left to right, the
grey ones walking in opposite direction. \\

\vspace*{9.5cm}
{\bf Fig. 2} ($N=500$, $S=500$, $p=1/2$): \\
In order to avoid collisions with other pedestrians the direction of walking
has to be changed often. \\

\clearpage
\thispagestyle{empty}
\vspace*{8.7cm}
{\bf Fig. 3} ($N=500$, $S=500$, $p=0.7$): \\
If the probability $p$
for passing a hindering pedestrian on the right is different from
the probability
$1-p$ for passing it on the left, two lanes of opposite direction develop. \\

\vspace*{10cm}
{\bf Fig. 4} ($N=350$, $S=540$, $p=0.7$): \\
In front of and behind an obstacle a pedestrian free area develops. \\

\clearpage
\thispagestyle{empty}
\vspace*{8.7cm}
{\bf Fig. 5} ($N=350$, $S=540$, $p=0.7$): \\
Obstacles at the margin of a pedestrian way reduce its effective width. \\
 
\clearpage
\thispagestyle{empty}
\vspace*{3cm}
\begin{center}
{\LARGE \bf A Mathematical Model for the Behavior \vspace{4mm}\\
of Pedestrians} \\
\vspace{0.5cm}
{\Large by \\
\vspace{0.5cm}
Dirk Helbing \\
II. Institut f\"{u}r Theoretische Physik \\
Universit\"{a}t Stuttgart \\
\vspace{2cm}
{\bf Abstract}}
\end{center}

{\raggedright The movement of pedestrians is} 
supposed to show certain regularities which can
be best described by an ``algorithm'' for the individual behavior and is easily
simulated on computers. This behavior is assumed to be determined by an
intended velocity, by several attractive and
repulsive effects and by fluctuations. 
The
movement of pedestrians is dependent on decisions, which have the purpose of
optimizing their behavior and can be explicitly modelled. Some interesting
applications of the model to real situations are given, especially to formation
of groups, behavior in queues, avoidance of collisions and selection
processes between behavioral alternatives.
\end{document}